\documentclass[journal]{IEEEtran}

\usepackage{graphicx}
\usepackage{cite}
\usepackage{float}
\usepackage{amsmath}
\usepackage{textcomp}
\usepackage{multirow}
\usepackage{xcolor}
\usepackage{hyperref}
\usepackage{mathtools, nccmath}
\usepackage[flushleft]{threeparttable} 
\usepackage{todonotes}
\usepackage[switch]{lineno}
\usepackage{graphics}
\usepackage{enumitem}

\newcommand{\revision}[1]{\textcolor{black}{#1}}

\begin{document}
\title{Automated Segmentation of Optical Coherence Tomography Angiography Images: Benchmark Data and Clinically Relevant Metrics}
%
%
 \author{Ylenia Giarratano,
 Eleonora Bianchi, Calum Gray, Andrew Morris, Tom MacGillivray, Baljean Dhillon and Miguel O. Bernabeu\thanks{Y. Giarratano, M. O. Bernabeu and A. Morris, are with the Usher Institute, University of Edinburgh, Edinburgh, EH16 4UX, UK. Emails: \{ylenia.giarratano, miguel.bernabeu, andrew.morris\}@ed.ac.uk}%
 \thanks{E. Bianchi is with Princess Alexandra Eye Pavilion, NHS Lothian, Edinburgh, EH3 9HA, UK. Email: e.bianchi@nhs.net}%
 \thanks{C. Gray is with Edinburgh Imaging, University of Edinburgh, Edinburgh, EH16 4TJ, UK. Email: calum.gray@ed.ac.uk}%
\thanks{A. Morris is with Health Data Research UK, London, NW1 2BE, UK.}%
\thanks{B. Dhillon is with the School of Clinical Sciences, University of Edinburgh, Edinburgh, EH3 9HA, UK. Email: baljean.dhillon@ed.ac.uk }%
\thanks{T. MacGillivray is with the Centre for Clinical Brain Sciences, University of Edinburgh, Edinburgh, EH16 4SB, UK. Email: t.j.macgillivray@ed.ac.uk }%
}



%
\maketitle              

\begin{abstract}
Optical coherence tomography angiography (OCTA) is a novel non-invasive imaging modality for the visualisation of microvasculature \textit{in vivo} that has encountered broad adoption in retinal research. OCTA potential in the assessment of pathological conditions and the reproducibility of studies relies on the quality of the image analysis. However, automated segmentation of parafoveal OCTA images is still an open problem. In this study, we generate the first open dataset of retinal parafoveal OCTA images with associated ground truth manual segmentations. Furthermore, we establish a standard for OCTA image segmentation by surveying a broad range of state-of-the-art vessel enhancement and binarisation procedures. We provide the most comprehensive comparison of these methods under a unified framework to date. Our results show that, for the set of images considered, \revision{deep learning architectures (U-Net and CS-Net)} achieve the best performance. For applications where manually segmented data is not available to retrain these approaches, our findings suggest that optimal oriented flux is the best handcrafted filter 
from those considered. Furthermore, we report on the importance of preserving network structure in the segmentation to enable deep vascular phenotyping. 
\revision{We introduce new metrics for network structure evaluation in segmented angiograms. Our results demonstrate that segmentation methods with equal Dice score perform very differently in terms of network structure preservation. 
Moreover, we compare the error in the computation of clinically relevant vascular network metrics (e.g. foveal avascular zone area and vessel density) across segmentation methods. Our results show up to 25\% differences in vessel density accuracy depending on the segmentation method employed. These findings should be taken into account when comparing the results of clinical studies and performing meta-analyses}. Finally, we release our data and source code to support standardisation efforts in OCTA image segmentation.

\end{abstract}

\section{Introduction}
\IEEEPARstart{A}{} number of studies have demonstrated that phenotypes of the retinal vasculature 
represent important biomarkers for early identification of pathological conditions such as diabetic retinopathy \cite{Jenkins2015}, cardiovascular disease \cite{Poplin2018}, and neurodegenerative disease \cite{DeBuc2017}.  Therefore, information regarding structural and functional changes in the retinal blood vessels can play a crucial role in the diagnosis and monitoring of these diseases. \\
Optical coherence tomography angiography (OCTA) is a novel non-invasive imaging modality that allows visualisation of the microvasculature \textit{in vivo} across retinal layers. It is based on the principle of repeating multiple OCT B-scans in rapid succession at each location on the retina. Static tissues will remain the same, while tissues containing flowing red blood cells will show intensity variations over time. OCTA can provide angiograms at different retinal depths and, unlike fluorescein angiography, does not require any dye injection, which may carry the risk of adverse reactions \cite{Musa2006}. \\
OCTA diagnosis potential has already been established in the context of neurovascular disease, diabetic retinopathy and, more recently, in chronic kidney disease. 
In \cite{Yoon2019}, microvascular characteristics calculated from OCTA images are compared between Alzheimer's disease patients, mild cognitive impairment (MCI) patients, and cognitive intact controls. Results showed a decrease in vessel density (VD) and perfusion density (PD)  of Alzheimer participants compared with the MCI and controls, opening to the possibility that changes in the retinal microvasculature may mirror small vessel disease in the brain, which is currently not possible to image clinically. 
Multiple studies on diabetic retinopathy have demonstrated that measurements  from the foveal avascular zone (FAZ), e.g., area and acircularity, in OCTA images are discriminant features in diabetic eyes compared to healthy
individuals, even before retinopathy develops \cite{Khadamy2018,Takase2015}. 
Finally, a recent study on renal impairment \cite{Vadala2019} demonstrated the potential of OCTA to find associations between changes in the retina and chronic kidney disease (CKD). OCTA scans revealed close association between CKD and lower paracentral retinal vascular density in hypertensive patients.\\
 Measurements used in these studies are based on quantifying phenotypes such as vessel density (VD), fractal dimension (FD), and percentage area of nonperfusion (PAN), extracted from binary masks of OCTA images \cite{Nesper2017, Reif2012}. However, the accuracy of these measurements and their reproducibility relies on the quality of the image segmentation. Since manual segmentation of blood vessels is a time consuming procedure that requires intra and inter-rater repeatability, there is a necessity to establish a fast automated method not affected by individual subjectivity.
The development of automated segmentation algorithms for OCTA images is a novel research field and no consensus exists in the literature about the best approaches. For example, in \cite{Alibhai2019} and \cite{Krawitz2017}, OCTA phenotypes are calculated on manually traced vessels. Simple thresholding procedures are used in \cite{Onishi2018,Nesper2017,Hwang2016}. Hessian filters followed by thresholding are applied on the original image to enhance vessels structure in \cite{Kim2016,Zhang2016}. A convolutional deep neural network approach was proposed in \cite{Prentasic2016} and more recently U-Net, and CS-Net architectures were adapted to OCTA in \cite{Mou2019}. However, how these different approaches compare to each other is not known. Furthermore, it is currently unknown how these methods perform when it comes to preserving network connectivity in the segmentation. This is a key aspect that can enable advanced vascular network phenotyping based on network science approaches (\emph{e.g.} \cite{AmatRoldan2015, Alves2018}).\\
 In this work, we take advantage of OCTA images from the PREVENT cohort (\url{https://preventdementia.co.uk/}), an ongoing prospective study aimed to predict early onset of dementia \cite{Ritchie2013}. \revision{Previous studies have shown OCTA imaging as a source of biomarkers for neurodegenerative disease \cite{Jiang2018, VanDeKreeke2019}, and together with MRI scans, OCTA images are being investigated in PREVENT.}
We derive and validate the first open dataset of retinal parafoveal OCTA images with associated ground truth manual segmentations. Furthermore, we establish a standard for OCTA image segmentation by surveying a broad range of state-of-the-art vessel enhancement and binarisation procedures. We provide the most comprehensive comparison of these methods under a unified framework to date. Furthermore, we report on the importance of preserving full network connectivity in the segmentation of angiograms to enable deep vascular phenotyping \revision{and introduce two new network structure evaluation metrics: the largest connected component ratio (LCC) and the topological similarity score (TopS). Our results show that, for the set of images considered, the U-Net and CS-Net architecture achieve the best performance in Dice score (both 0.89), but the latter reaches a better performance in TopS. Among the handcrafted filter enhancement methods from those considered, optimal oriented flux is the best in both pixelwise and network metrics. 
Our results demonstrate that methods with equal Dice score (\emph{e.g.} adaptive thresholding and OOF) can perform substantially different in terms of LCC or TopS. 
Furthermore, we compare the relative error in the computation of clinically relevant vascular network metrics (\emph{e.g.} foveal avascular zone area and vessel density) across segmentation methods. 
Our results show up to 25\% differences in vessel density and 24\% in FAZ area depending on the method employed and that U-Net outperfoms all other methods when investigating the FAZ. These findings should be considered when comparing the results of clinical studies and performing meta-analyses.} Finally, we release our data and source code to support standardisation efforts in OCTA image segmentation.

 
\section{Methods}
\subsection{Data acquisition and Manual segmentation}
Imaging  was performed using the commercial RTVue-XR Avanti OCT system (OptoVue, Fremont, CA). Consequent B-scans, each one consisting of 304$\times$304 A-scans, were generated in $3\times3$ mm and $6\times6$ mm fields of view centered at the fovea. Maximum decorrelation value is used to generate \textit{en face} angiograms of superficial, deep and choriocapillary layers.
In this work, we selected images only of the superficial layer (containing the vasculature enclosed in the internal limiting membrane layer (ILM) and the inner plexiform layer (IPL)) with 3$\times$3 mm field of view from left and right eyes of participants with and without family history of dementia as part of a prospective study aimed to find early biomarkers of neurodegenerative diseases (PREVENT). 
From the initial 17 participants we extracted five subimages, one from each clinical region of interest (ROI): superior, nasal, inferior, temporal, and foveal (Figure \ref{ROIS}A), and a target of 55 ROIs among the best quality images was set for the purpose of this study.  Criteria for the exclusion were based on major visible artifacts \cite{Spaide}. The final 55 ROIs were selected from 11 participants  and split into training (30 ROIs) and test (25 ROIs).

\subsection{Manual segmentation}
\begin{figure*}[!htb]
\includegraphics[width=\textwidth]{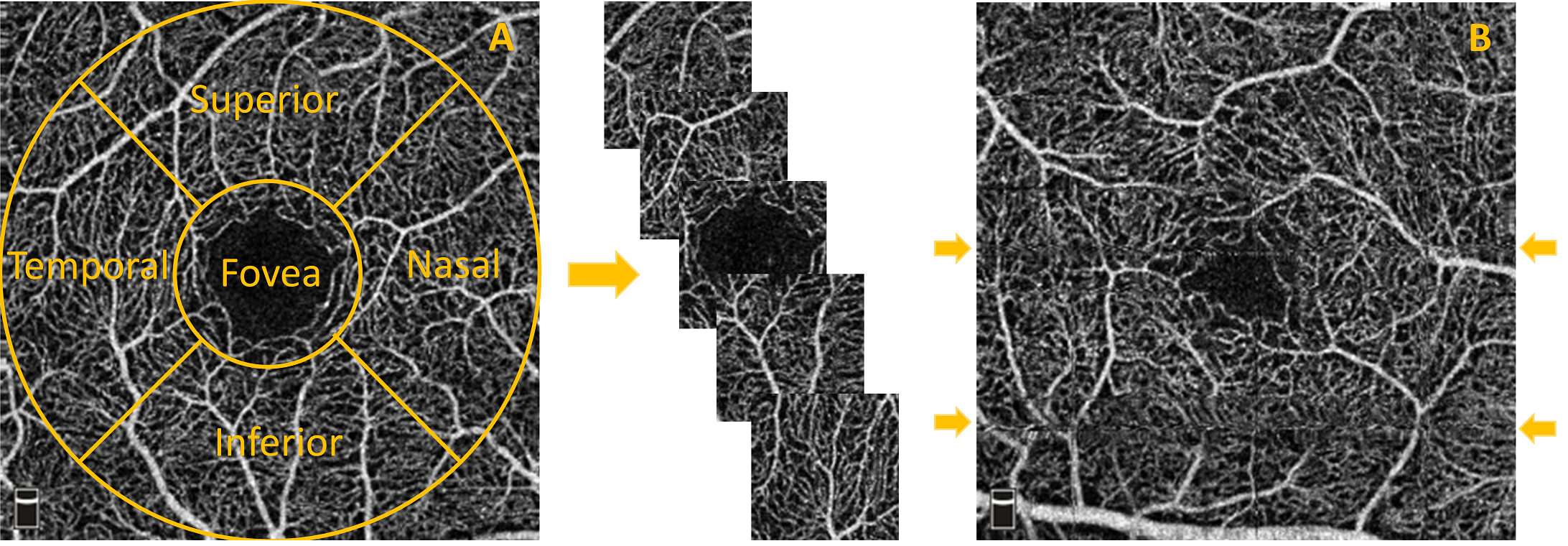}
\caption{(A) Extraction of images from each clinical region of interest: superior, nasal, foveal, inferior, and temporal. (B) Examples (arrows) of horizontal artifacts in OCTA images.} \label{ROIS}
\end{figure*}

A number of challenges need to be overcome in OCTA manual segmentation: images suffer from poor contrast, low signal to noise ratio and can contain motion artifacts generated during the scan acquisition. The most common visible artifacts are vertical and horizontal line distortions, as shown in Figure \ref{ROIS}B. 
Furthermore, the fact that images are constructed from the average of a volume means that, in our segmentation, we cannot distinguish vessels going past each other at different depths. In general, bigger vessels appear brighter and easier to trace, however, the smallest capillaries are challenging to segment and therefore are affected to subjective interpretation by any given rater. \\
Previous OCTA studies have performed manual continuous blood vessel delineation with or without consideration of vessel width (\cite{Prentasic2016} and \cite{Mou2019}, respectively). Given the sources of uncertainty previously described, this approach may overinterpret vessel connectivity and suffer from reproducibility issues that remain currently unexplored in the literature. Instead, we adopted a more conservative approach and performed pixelwise manual segmentation selecting all pixels enclosed in the vasculature (using the ITK-SNAP software \cite{ITK-SNAP}). A previous study performing pixelwise segmentation \cite{Eladawi2017} did not assess reproducibility of the segmentations and could not resolve the finest capillaries in the scans. 

\subsection{Automated image segmentation methods}
Vessel enhancement approaches consist of filters that improve the contrast between vessels and background. We chose four well-known handcrafted filters for blood vessel segmentation, based on implementation availability and previous applications to the enhancement of tubular-like structures in retinal images: \textit{Frangi} \cite{Frangi1998}, 
\textit{Gabor} \cite{Soares2006},
\textit{SCIRD-TS} \cite{Annunziata2016}, 
and \textit{OOF} \cite{Law2008}. All these filters require parameter tuning. In our case, from a range of possible configurations we selected the optimal set of parameters that gave the best performance when compared to the manual segmentation (see Table \ref{tab: params}).\\ 
Although handcrafted filters work in many cases, often real images do not satisfy their assumptions (\textit{e.g.} locally tubular structure  and gaussian intensity profile). To overcome this issue probabilistic  and machine learning frameworks have been proposed \cite{Eladawi2017}, \cite{Prentasic2016}. In this study, we considered the latter by adopting three deep learning architectures. We used a pixelwise convolutional neural network (CNN), U-Net, and the more recently proposed CS-Net \cite{Mou2019}. 
The design of the CNN for pixelwise classification is based on the one proposed in \cite{Prentasic2016} for OCTA segmentation.
It consists of three convolutional layers with rectified linear unit activation (ReLU), each followed by maxpooling. To reduce the risk of overfitting, dropout is used before the last  fully connected layer. Cross entropy and adam optimizer were used during the learning process. For each training image we randomly extracted the same number of vessel and background pixels to balance the classes. A patch containing the pixel to classify and its $61\times61$ neighbourhood is used as input to the network. More than 200,000 patches were used during the training. Finally, the probability of belonging to a vessel or background is then used to generate the enhanced grayscale image (see Table \ref{tab:CNN}). \\
Developed for biomedical image segmentation, U-Net is a fully convolutional neural network characterised by a contracting path and an expansive path that confer to the network its U shape. It has proved to be fast and accurate even with few training images. The architecture consists of modules of two repeated convolutional layers with ReLU activation function followed by maxpooling for the encoder path, upsampling and two repeated convolutional layers for the decoder path (see Table \ref{tab:Unet}). The lowest resolution is $8\times8$ pixels, binary cross entropy is used as loss function, and SGD as optimizer. From each ROI, 1,000 patches of size $32\times32$ are extracted to train the network, for a total of 30,000 training inputs.\\ Finally, the recently proposed CS-Net was tested using the same sub-patches procedure previously described.  As U-Net, this architecture is characterised by a contractive and a expansive path. However between those paths, two other elements are present: the spatial and channel attention module. The first uses spatial correlation to acquire global contextual features, the latter uses changes in intensity across channels to extract features. Adam optimizer and MSE loss are used to train the model. Given our initial sample size, data augmentation (flipping horizontally  or vertically) has been used in all the three architectures with the 10\% of training inputs used as validation set.\\
Vessel enhancement is often followed by a threshold step to obtain the vessel binary mask. However, modern methods employ the enhanced vasculature as a preliminary step for more advanced binarisation algorithms, such as machine learning (ML) classifiers. 
In this work we use adaptive thresholding as baseline binarization procedure, a method that 
takes into account spatial variations in illumination \cite{Bradley} in a specified neighbourhood of the pixel. We compared this approach with other binarisation approaches,
support vector machines (SVMs), random forest (RF) and k-nearest neighbours (k-NN) as binarisation procedure in the case of Frangi, Gabor, SCIRD-TS. A two step binarisation procedure, suggested in \cite{Li2017}, has been used in the case of OOF, a global threshold for larger vessels and adaptive threshold for the smallest ones. Finally, a global thresholding, based on the shape of the pixel intensity histogram, is used to binarise the probability maps obtained from the CNN architecture, adaptive thresholding is applied to the output of U-Net, and the Otsu method \cite{Otsu} in the case of CS-Net. After all binarisation procedures morphological opening is performed to remove small disconnected pixel structures. 
In each of the ML binary classifiers we used seven features to characterise pixels: intensity based features extracted from a $3\times3$ pixel neighbourhood (intensity value, range, average, standard deviation, and entropy) and geometric features (the local curvature information provided by the hessian eigenvalues), \cite{Rodrigues2013}. 

\subsection{Segmentation evaluation}
Cohen's kappa coefficient is a robust statistic for testing intra and inter-rater variability \cite{Kappa}. 
Considering $Pr(a)$ and $Pr(e)$ as the observed agreement and the chance agreement, respectively, it can be computed as:
\begin{equation}\small
    \kappa = \frac{Pr(a)-Pr(e)}{1-Pr(e)}.
\end{equation}
In our study $Pr(a)$ is the accuracy in pixel classification (vessel \textit{vs} background)  and $Pr(e)$ is the sum of the probability of both raters randomly selecting vessel pixels and the probability of both of them randomly selecting background pixels for a given ROI. \\ For the ROIs in the test set,
pixelwise comparison between manual and automated segmentation was performed using accuracy, precision, recall along with  Dice similarity coefficient defined as
\begin{equation}\small
Dice = \frac{2TP}{2TP+FP+FN},
\end{equation}
where TP, FP, FN represent true positive, false positive and false negative, respectively.\\
Furthermore, for the evaluation of the \revision{global quality of segmentation,} 
we used the CAL metric proposed in \cite{Gegundez-Arias2012}. It is based on three descriptive features:
\begin{itemize}[leftmargin=*]
    \item connectivity (C), to assess the fragmentation degree between segmentations, described mathematically by the formula
    \begin{equation} C(S, S_{GT}) = 1 - min \left ( 1,\frac{ |\#_CS_{GT}- \#_C S|}{\#S_{GT}} \right) \end{equation}
    where $\#_CS$ and $\#_CS_{GT}$ are the number of connected components in the segmented and ground truth image, while $\#S_{GT}$ is the number of vessel pixels in the mask;
    \item area (A), to evaluate the degree of overlapping, defined as 
        \begin{equation} A(S, S_{GT}) = \frac{\#((\delta_{\alpha}(S) \cap S_{GT} ) \cup (S \cap \delta_{\alpha}(S_{GT}))}{\#(S \cup S_{GT})}, 
        \end{equation}
     where $\delta_{\alpha}$ is a morphological dilatation using a disc of radius $\alpha$;
    \item length (L), to capture the degree of coincidence, described by 
        \begin{equation} L(S, S_{GT}) = \frac{\#((\varphi(S) \cap \delta_{\beta} (S_{GT})) \cup (\delta_{\beta}(S) \cap \varphi(S_{GT}))}{\#(\varphi(S) \cup \varphi(S_{GT}))}, \end{equation}
     where $\varphi$ indicates a skeletonisation procedure and $\delta_{\beta}$ is a morphological dilatation using a disc of radius $\beta$.
\end{itemize}
Considering vessel width and closeness between capillaries in OCTA images, we set $\alpha$ and $\beta$ both equal to 1. The product of C, A, and L, (CAL) results sensitive to the vascular features and takes values in the range [0,1], with the zero denoting the worst segmentation and 1 the perfect segmentation.\\
\revision{Despite CAL contains a connectivity component, its effect is weighted by the values of the area (A) and length (L) metrics. 
Hence  we introduce a new metric, namely the largest connected component ratio (LCC), with the aim of penalising those methods that do not retrieve connections of the vascular network. LLC is defined as:}
\begin{equation}
    LCC = 1 - min(1, \frac{\left|\# LCC_S - \# LCC_{GT}\right|}{\#LCC_{GT}}), 
\end{equation}
where $\#LCC_{S}$ and $\#LCC_{GT}$ are the lengths, in terms of number of pixels, of the largest connected component in the skeleton of the segmented and ground truth images.
The closest to 1 is the LCC ratio, the more similar is in structure the largest connected component of the segmented image compared to the ground truth. \revision{Using LCC in conjunction with Dice score, we provide information about  pixelwise similarity and connectivity. Indeed, a single pixel difference does not affect Dice, however it may drastically change the number of connected components, affecting the LCC ratio.}\\
 \revision{
To evaluate the topological accuracy of a segmentation, we use the concept of persistent homology and Betti numbers for angiograms described in \cite{Hu2019a}. By introducing the topological similarity score (TopS) defined as 
\begin{equation}
    TopS = 1 - min(1, \frac{\left| {\beta_1}{_S} -  {\beta_1}_{GT}\right|}{{\beta_1}_{GT}}), 
\end{equation}
where $\beta_{1}$ is the first Betti number associated with the image, and indicating the counts of one-dimensional holes, we compute the similarity between topological structures. \\
Finally, popular biomarkers in the literature for OCTA images include vessel density, FAZ area, and FAZ acircularity index. We investigate how the segmentation method affects these metrics by reporting the relative error against ground truth measurements.
Vessel density is defined as the number of white pixels over the total number of pixels. To compute FAZ area and acircularity index, we convert the skeleton of the image into a graph object where the largest loop (face) in the network is identified as the FAZ \cite{Schneider}. This method takes into account a continuous boundary, therefore images with disconnected contour will show greater FAZ area.}


\section{Results}
\subsection{Inter and intra-rater agreements}
The ground truth dataset contains 55 ROIs segmented by one rater (rater A). \revision{Rater A (Y.G.) segmented 20 images twice to assess the intra rater agreement. Another set of 20 images was segmented by rater B (E.B.) to determine the inter-rater reliability.}
Results show good agreements (for each pair $k> 0.7$) with an average of $0.8$ for the intra rater agreement and $0.77$ between operators, which demonstrates that the proposed approach to segmentation is reproducible.

\subsection{Automated approaches for pixelwise classification}
\begin{figure*}[!htb]
\centering
\includegraphics[width=14cm]{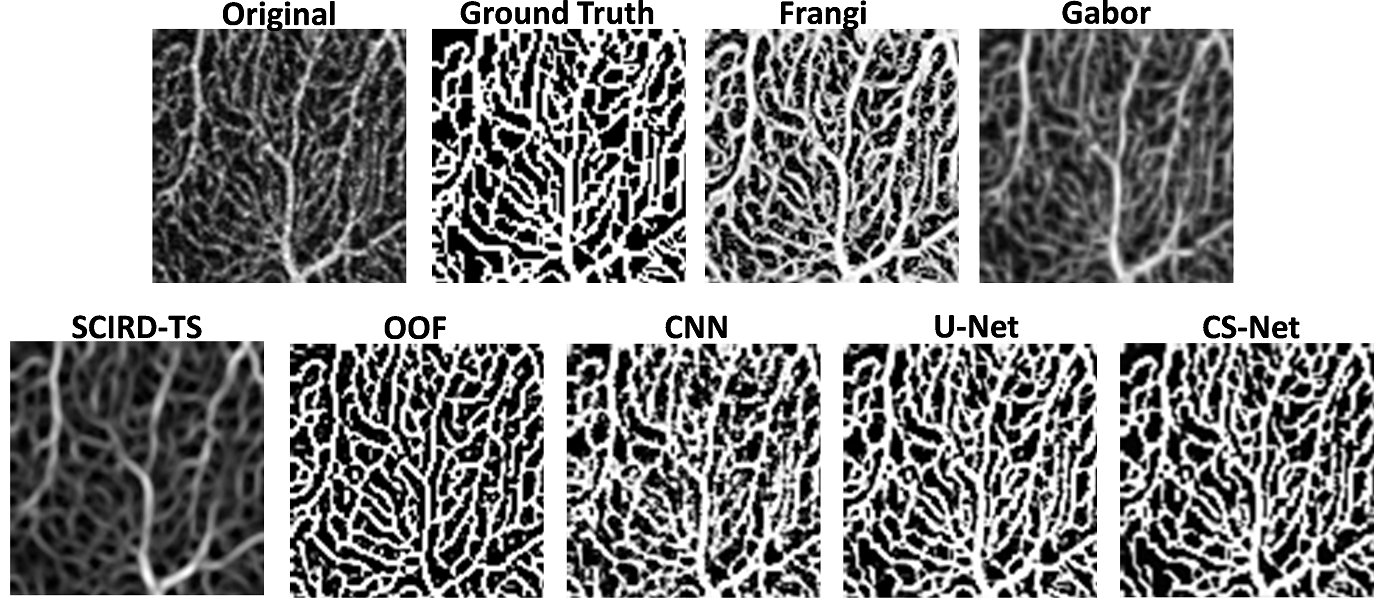}
\caption{Example of vessels enhancement. Original, ground truth and images after vessel enhancement by using Frangi, Gabor, SCIRD-TS, OOF, CNN, U-Net, CS-Net.} \label{enhancers}
\end{figure*}

\begin{figure*}[!htb]
\centering
\includegraphics[width=14cm]{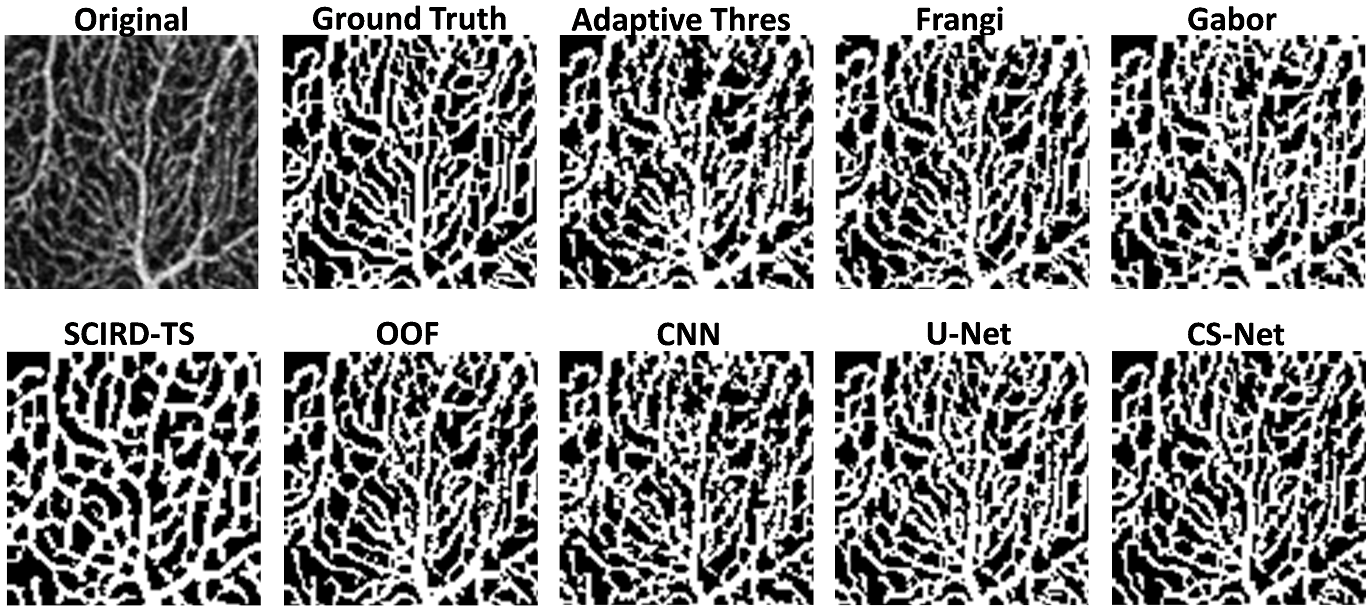}
\caption{Vessel segmentation in superior parafoveal OCTA image. Original, ground truth images followed by binary images after vessel enhancement by using Frangi (+RF), Gabor (+RF), SCIRD-TS (+SVM), OOF, CNN, U-Net, and CS-Net.} \label{segmentations}
\end{figure*}

\begin{figure*}[!htb]
\centering
\includegraphics[width=\textwidth]{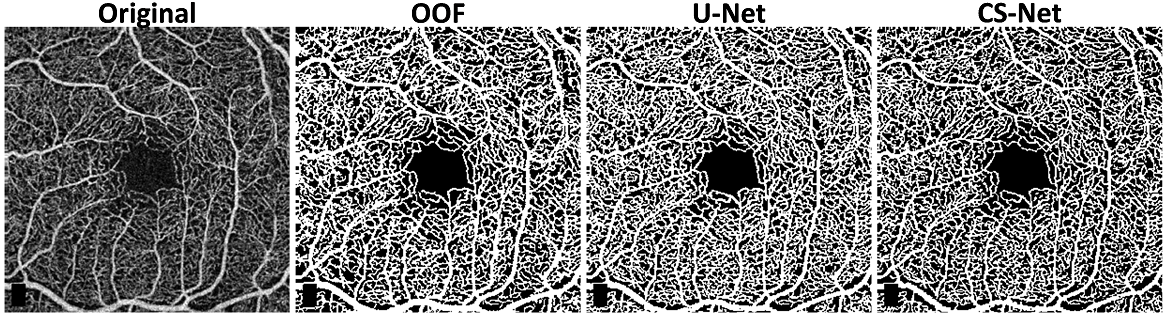}
\caption{Whole image segmentation by using the best three methods, OOF, U-Net, and CS-Net. Optovue RTVue XR Avanti scan logo on the bottom left corner was removed from the original image.} \label{whole_IM}
\end{figure*}

Segmentation performances according to the metrics proposed are shown in Table \ref{tab:table1}. U-Net \revision{and CS-Net} outperform all the other methods, by reaching a Dice score of 0.89. Among the handcrafted filters, OOF and Frangi filters achieve good performances with an average Dice score of 0.86 and 0.85, respectively. \revision{Our baseline method, adaptive thresholding without vessel enhancement, achieves comparable Dice performance. However, it encounters difficulties resolving network connectivity as shown by the LCC and TopS metrics compared to Frangi and OOF.} The use of machine learning methods as binarisation procedure can improve performance compared to thresholding after Frangi, Gabor and SCIRD-TS both in terms of pixelwise and network structure accuracy.  
\revision{Deep learning architectures reach the best results in LCC ratio (CNN, 0.94, UNet and CS-Net, 0.93) together with the OOF (0.94). The highest TopS score is reached by the CS-Net and OOF, 0.83 and 0.80, respectively. Moreover, the same two methodologies achieve the two lowest vessel density error (6\%, and 10\%).} 
Investigating the enhanced images (Figure \ref{enhancers}), we noticed that each method suffers from different deficiencies. Frangi filter clusters nearby vessels, losing important information contained in the microvasculature. Gabor filter enhances centrelines, performing poorly on the detection of vessel edges. SCIRD-TS remodels the vasculature making it more regular and equally spaced. OOF retrieves the smallest capillaries but overenhances noise in the foveal region. Figure \ref{segmentations} shows segmentation results after applying each vessel enhancement method and best binarisation procedure.
Figure \ref{whole_IM} shows whole image segmentations with the best handcrafted and learnt filters.

\begin{table}[h!]

    \caption{Segmentation performance \revision{(best method per column in bold )}}
    \label{tab:table1}
    \resizebox{9cm}{!}{%
    \begin{tabular}{c|cccccccc} 
        \textbf{Method} &  \textbf{Dice}&  \revision{\textbf{Acc}}&  \revision{\textbf{Rec}}&  \revision{\textbf{Pre}} &  \textbf{CAL} &  \textbf{LCC} &  \revision{\textbf{TopS}}& \revision{\textbf{VD}}  \\ \hline
        \revision{ \textbf{Adaptive thres (AT)}} &  \revision{0.86} &\revision{ 0.89} &\revision{\textbf{0.89}} & \revision{0.92} & 0.83 &  0.83 & \revision{0.70} &\revision{14\%} \\
        \\
      \textbf{Frangi+AT} &  0.83 &\revision{ 0.86} &\revision{0.83} & \revision{0.93} & 0.83 &  0.88 & \revision{0.72} &\revision{21\%} \\
       \textbf{Gabor+AT} & 0.77 & \revision{0.81} & \revision{0.78} & \revision{0.87} & 0.75 & 0.76 & \revision{0.59} & \revision{24\%} \\
        \textbf{SCIRD-TS+AT} & 0.71 & \revision{0.76} & \revision{0.74} & \revision{0.82} & 0.66 & 0.68 & \revision{0.46} &\revision{25\%} \\
                 \textbf{OOF} & 0.86 & \revision{0.88} & \revision{0.87} & \revision{0.92} & 0.85 & \textbf{0.94} & \revision{0.80} & \revision{10\%} \\
        \\
          \textbf{Frangi+k-NN} & 0.84 & \revision{0.87} & \revision{0.85} & \revision{0.91} & 0.86 & 0.91 & \revision{0.59} &\revision{14\%}\\
      \textbf{Frangi+SVM} & 0.85 & \revision{0.88} & \revision{0.85} & \revision{0.93} &0.87 &  0.94 & \revision{0.76} &\revision{15\%} \\
      \textbf{Frangi+RF} & 0.85 & \revision{0.88} & \revision{0.86} & \revision{0.92} & 0.87 &  0.94 & \revision{0.75} & \revision{13\%} \\
      \textbf{Gabor+k-NN} & 0.82 & \revision{0.84} & \revision{0.80} & \revision{0.92} & 0.84 & 0.84 & \revision{0.37} & \revision{21\%}  \\
      \textbf{Gabor+SVM} & 0.83 & \revision{0.85} & \revision{0.78} & \revision{0.94} & 0.85 & 0.84 & \revision{0.45} &\revision{24\%} \\
      \textbf{Gabor+RF} & 0.83 & \revision{0.85} & \revision{0.80} & \revision{0.93} & 0.85 & 0.87 & \revision{0.45} &\revision{22\%}\\

      \textbf{SCIRD-TS+k-NN} & 0.72 & \revision{0.77} & \revision{0.76} & \revision{0.82} &  0.74 & 0.90 & \revision{0.35} &\revision{19\%} \\
      \textbf{SCIRD-TS+SVM} & 0.75 & \revision{0.79} & \revision{0.78} & \revision{0.84} &  0.75 & 0.75 &  \revision{0.54} &\revision{19\%} \\
      \textbf{SCIRD-TS+RF} & 0.74 & \revision{0.78} & \revision{0.77} & \revision{0.83} & 0.75 & 0.80 & \revision{0.65} &\revision{19\%}\\
             \\
 
      \textbf{CNN} & 0.83 & \revision{0.86} & \revision{0.85} & \revision{0.91}  & 0.85 & \textbf{0.94} & \revision{0.70} & \revision{14\%} \\
      \textbf{U-Net} & \textbf{0.89} & \revision{\textbf{0.91}} & \revision{0.87} & \revision{\textbf{0.97}} &  \textbf{0.90} & 0.93& \revision{0.67} &\revision{17\%} \\
      \revision{\textbf{CS-Net}} & \revision{\textbf{0.89}} & \revision{\textbf{0.91}} & \revision{\textbf{0.91}} & \revision{0.93} & \revision{\textbf{0.90}} & 0.93 & \revision{\textbf{0.83}} & \revision{\textbf{6\%}} \\

    \end{tabular}%
    }
\end{table}

\subsection{Foveal avascular zone}
Foveal images are characterised by the presence of a predominantly dark area free from blood vessels called foveal avascular zone (FAZ). We noticed that handcrafted filters have difficulties with those images, overenhancing noise in the central region ( Fig. \ref{fovea}(A)). This can lead to vessel detection in the FAZ when segmentation by simple thresholding is applied. \revision{Machine learning methods were less affected by this issue  since they learnt from the ground truth data. 
Motivated by this finding, we investigated segmentation performance on the 5 different ROIs. Table \ref{tab:table2} shows that, except for CS-Net, the foveal region has consistently the lowest Dice score across segmentation methods. 
Visual inspection of our segmentations (see  Fig. \ref{fovea}(B)-(C) reveal that incorrect detection of the boundary of the FAZ leads to important errors in FAZ area (FazE) and acircularity index (AIE) (see Table \ref{tab:table2}). 
}

\begin{table}[h!]
\begin{center}
\caption{ \revision{Dice score per ROI (foveal (F), superior (S), nasal (N), inferior (I), temporal (T), and foveal (F)) and FAZ error metrics}}
\label{tab:table2}
  \resizebox{\columnwidth}{!}{%
\begin{tabular}{c|ccccc|cc} 
         \revision{\textbf{Method}} &  \revision{\textbf{F}}&  \revision{\textbf{S}}&  \revision{\textbf{N}}&  \revision{\textbf{I}} &   \revision{\textbf{T}} & 
         \revision{\textbf{FazE }} &
         \revision{\textbf{AIE}}\\    \hline
 
   \revision{Adaptive thres} &   \revision{0.84} &  \revision{0.87}&  \revision{0.88}&  \revision{0.88} &   \revision{0.86} & \revision{14\%}& \revision{5\%} \\   
    
  \revision{Frangi + thres} &   \revision{0.79} &  \revision{0.86}&  \revision{0.86}&  \revision{0.85} &   \revision{0.85}  &\revision{12\%}& \revision{9\%}\\
                                
 \revision{Gabor + thres} &   \revision{0.75}&  \revision{0.78}&  \revision{0.80}&  \revision{0.75} &   \revision{0.77}  &\revision{13\%}& \revision{10\%}\\
                
\revision{SCIRD-TS + thres} &  \revision{0.71}&  \revision{0.73}&  \revision{0.75}&  \revision{0.67} &  \revision{0.71} &\revision{14\%}&\revision{7\%} \\
                                 
\revision{OOF} &  \revision{0.84}&  \revision{0.86}&  \revision{0.88}&  \revision{0.86} &  \revision{0.85} &\revision{24\%}& \revision{11\%}\\
                                                 
\revision{CNN} &  \revision{0.82}&  \revision{0.84}&  \revision{0.85}&  \revision{0.84} &  \revision{0.83} &\revision{6\%}& \revision{6\%}\\
                                
\revision{U-Net} &  \revision{0.87}&  \revision{0.90}&  \revision{0.90}&  \revision{0.90} &  \revision{0.89} &\revision{\textbf{5\%}} & \revision{\textbf{4\%}} \\

\revision{CS-Net} &  \revision{0.89}&  \revision{0.89}&  \revision{0.90}&  \revision{0.89} &  \revision{0.88} & \revision{14\%}& \revision{5\%} \\
                            \end{tabular}%
                            }
  \end{center}
\end{table}
\section{Discussion and Conclusions}
Retinal image analysis has demonstrated great potential for the discovery of biomarkers of eye and systemic disease. Recently, OCTA imaging has enabled the visualisation of the smallest capillaries in the retina without the need of a contrast agent. However, its potential for the assessment of pathological conditions and the reproducibility of studies based on it relies on the quality of the image analysis. Automated OCTA image segmentation is an open problem in the field. In this study, we generate the first open dataset of retinal parafoveal OCTA images with associated ground truth manual segmentations. We pay special attention to segmenting the images in a reproducible way and demonstrate good inter- and intra-rater agreement. 
We present a comparison of  state-of-the-art vessel enhancement and binarisation procedures under a unified computational framework and make the source code available. 
By introducing two novel metrics, we evaluate segmentation quality measures to guide the identification of the algorithm that not only provides the best agreement with the manually segmented images, but also achieves the best preservation of their network structure. \\ 
\revision{Our study shows that CS-Net reaches the best performances in almost all the considered evaluation metrics, suggesting this method as the best segmentation approach for parafoveal OCTA image segmentation. Interestingly, OOF achieves segmentation performances in line with the neural network architectures without the requirement of extensive manually segmented images for training purposes. 
Our results highlight challenges in the segmentation of the FAZ: handcrafted filters suffer from noise enhancement in this region, indicating the necessity of masking that area or the use of denoising preprocessing procedures and more sophisticated binarisation methods when those filters are applied. 
Moreover our study underlines how clinically relevant metrics used to analyse OCTA images are sensitive to the segmentation method. 
Results show up to 25\% differences in vessel density accuracy depending on the method employed, with differences up to 11\% for methods with identical results in terms of pixelwise segmentation Dice and accuracy, suggesting that precaution should be taken when comparing the results of clinical studies and performing meta-analyses.} 

\section*{Source code and data availability}
OCTA images and rater A (Y.G.) segmentations are available in \url{https://doi.org/10.7488/ds/2729}. Handcrafted filter code was implemented in MATLAB R2018b (Version 9.5). Python 3.6.9 was used to build ML methods. Keras library with Tensorflow backend was used to implement the CNN and U-Net, \revision{and Pytorch for CS-Net. Gudhi library was used to compute the topological metric\cite{gudhi:urm}}. Source code available at \url{https://github.com/giaylenia/OCTA_segm_study}. 

\section*{Acknowledgements}
YG is supported by the Medical Research Council (MRC). MOB is supported by grants from EPSRC (EP/R029598/1, EP/R021600/1, EP/T008806/1), Fondation Leducq (17 CVD 03), and the European Union’s Horizon 2020 research and innovation programme under grant agreement No 801423. We thank the PREVENT research team and study participants. Image acquisition was carried out at the Edinburgh Imaging facility QMRI, University of Edinburgh. The research team acknowledges the financial support of NHS Research Scotland (NRS), through Edinburgh Clinical Research Facility.
The authors would like to thank the VAMPIRE team (\url{https://vampire.computing.dundee.ac.uk}) for fruitful discussions.


%
%
%
\vspace{-0.3cm}
\medskip

\bibliographystyle{IEEEtran}
\bibliography{MyCollection.bib}{}

\begin{thebibliography}{10}
\providecommand{\url}[1]{#1}
\csname url@samestyle\endcsname
\providecommand{\newblock}{\relax}
\providecommand{\bibinfo}[2]{#2}
\providecommand{\BIBentrySTDinterwordspacing}{\spaceskip=0pt\relax}
\providecommand{\BIBentryALTinterwordstretchfactor}{4}
\providecommand{\BIBentryALTinterwordspacing}{\spaceskip=\fontdimen2\font plus
\BIBentryALTinterwordstretchfactor\fontdimen3\font minus
  \fontdimen4\font\relax}
\providecommand{\BIBforeignlanguage}[2]{{%
\expandafter\ifx\csname l@#1\endcsname\relax
\typeout{** WARNING: IEEEtran.bst: No hyphenation pattern has been}%
\typeout{** loaded for the language `#1'. Using the pattern for}%
\typeout{** the default language instead.}%
\else
\language=\csname l@#1\endcsname
\fi
#2}}
\providecommand{\BIBdecl}{\relax}
\BIBdecl

\bibitem{Jenkins2015}
A.~J. Jenkins, M.~V. Joglekar, A.~A. Hardikar, A.~C. Keech, D.~N. O'Neal, and
  A.~S. Januszewski, ``{Biomarkers in diabetic retinopathy},'' \emph{Review of
  Diabetic Studies}, vol.~12, no. 1-2, pp. 159--195, 2015.

\bibitem{Poplin2018}
R.~Poplin, A.~V. Varadarajan, K.~Blumer, Y.~Liu, M.~V. McConnell, G.~S.
  Corrado, L.~Peng, and D.~R. Webster, ``{Prediction of cardiovascular risk
  factors from retinal fundus photographs via deep learning},'' \emph{Nature
  Biomedical Engineering}, vol.~2, no.~3, pp. 158--164, 2018.

\bibitem{DeBuc2017}
D.~C. DeBuc, G.~M. Somfai, and A.~Koller, ``{Retinal microvascular network
  alterations: Potential biomarkers of cerebrovascular and neural diseases},''
  \emph{American Journal of Physiology - Heart and Circulatory Physiology},
  vol. 312, no.~2, pp. H201--H212, 2017.

\bibitem{Musa2006}
F.~Musa, W.~J. Muen, R.~Hancock, and D.~Clark, ``{Adverse effects of
  fluorescein angiography in hypertensive and elderly patients},'' \emph{Acta
  Ophthalmologica Scandinavica}, vol.~84, no.~6, pp. 740--742, 2006.

\bibitem{Yoon2019}
S.~P. Yoon, D.~S. Grewal, A.~C. Thompson, B.~W. Polascik, C.~Dunn, J.~R. Burke,
  and S.~Fekrat, ``{Retinal Microvascular and Neurodegenerative Changes in
  Alzheimer's Disease and Mild Cognitive Impairment Compared with Control
  Participants},'' \emph{Ophthalmology Retina}, vol.~3, no.~6, pp. 489--499,
  2019.

\bibitem{Khadamy2018}
J.~Khadamy, K.~Aghdam, and K.~Falavarjani, ``An update on optical coherence
  tomography angiography in diabetic retinopathy,'' \emph{Journal of Ophthalmic
  \& Vision Research}, vol.~13, p. 487, 10 2018.

\bibitem{Takase2015}
N.~Takase, M.~Nozaki, A.~Kato, H.~Ozeki, M.~Yoshida, and Y.~Ogura,
  ``{Enlargement of foveal avascular zone in diabetic eyes evaluated by en face
  optical coherence tomography angiography},'' \emph{Retina}, vol.~35, no.~11,
  pp. 2377--2383, 2015.

\bibitem{Vadala2019}
M.~Vadal{\`{a}}, M.~Castellucci, G.~Guarrasi, M.~Terrasi, T.~{La Blasca}, and
  G.~Mul{\`{e}}, ``{Retinal and choroidal vasculature changes associated with
  chronic kidney disease},'' \emph{Graefe's Archive for Clinical and
  Experimental Ophthalmology}, vol. 257, no.~8, pp. 1687--1698, 2019.

\bibitem{Nesper2017}
P.~L. Nesper, P.~K. Roberts, A.~C. Onishi, H.~Chai, L.~Liu, L.~M. Jampol, and
  A.~A. Fawzi, ``{Quantifying Microvascular Abnormalities With Increasing
  Severity of Diabetic Retinopathy Using Optical Coherence Tomography
  Angiography},'' \emph{Investigative ophthalmology {\&} visual science},
  vol.~58, no.~6, pp. BIO307--BIO315, 2017.

\bibitem{Reif2012}
R.~Reif, J.~Qin, L.~An, Z.~Zhi, S.~Dziennis, and R.~Wang, ``{Quantifying
  optical microangiography images obtained from a spectral domain optical
  coherence tomography system},'' \emph{International Journal of Biomedical
  Imaging}, vol. 2012, 2012.

\bibitem{Alibhai2019}
A.~Y. Alibhai, E.~M. Moult, R.~Shahzad, C.~B. Rebhun, C.~Moreira-neto,
  M.~Mcgowan, D.~Lee, B.~Lee, C.~R. Baumal, A.~J. Witkin, E.~Reichel, J.~S.
  Duker, J.~G. Fujimoto, and N.~K. Waheed, ``{HHS Public Access},'' vol.~2,
  no.~5, pp. 418--427, 2019.

\bibitem{Krawitz2017}
B.~D. Krawitz, S.~Mo, L.~S. Geyman, S.~A. Agemy, N.~K. Scripsema, P.~M. Garcia,
  T.~Y. Chui, and R.~B. Rosen, ``{Acircularity index and axis ratio of the
  foveal avascular zone in diabetic eyes and healthy controls measured by
  optical coherence tomography angiography},'' \emph{Vision Research}, vol.
  139, pp. 177--186, 2017.

\bibitem{Onishi2018}
A.~C. Onishi, P.~L. Nesper, P.~K. Roberts, G.~A. Moharram, H.~Chai, L.~Liu,
  L.~M. Jampol, and A.~A. Fawzi, ``{Importance of considering the middle
  capillary plexus on OCT angiography in diabetic retinopathy},''
  \emph{Investigative Ophthalmology and Visual Science}, vol.~59, no.~5, pp.
  2167--2176, 2018.

\bibitem{Hwang2016}
T.~S. Hwang, S.~S. Gao, L.~Liu, A.~K. Lauer, C.~J. Flaxel, D.~J. Wilson,
  D.~Huang, Y.~Jia, and O.~Health, ``{HHS Public Access},'' vol. 134, no.~4,
  pp. 367--373, 2016.

\bibitem{Kim2016}
A.~Y. Kim, Z.~Chu, A.~Shahidzadeh, R.~K. Wang, C.~A. Puliafito, and A.~H.
  Kashani, ``{Quantifying microvascular density and morphology in diabetic
  retinopathy using spectral-domain optical coherence tomography
  angiography},'' \emph{Investigative Ophthalmology and Visual Science},
  vol.~57, no.~9, pp. OCT362--OCT370, 2016.

\bibitem{Zhang2016}
M.~Zhang, T.~S. Hwang, C.~Dongye, D.~J. Wilson, D.~Huang, and Y.~Jia,
  ``{Automated quantification of nonperfusion in three retinal plexuses using
  projection-resolved optical coherence tomography angiography in diabetic
  retinopathy},'' \emph{Investigative Ophthalmology and Visual Science},
  vol.~57, no.~13, pp. 5101--5106, 2016.

\bibitem{Prentasic2016}
P.~Prenta{\v{s}}ic, M.~Heisler, Z.~Mammo, S.~Lee, A.~Merkur, E.~Navajas, M.~F.
  Beg, M.~{\v{S}}arunic, and S.~Loncaric, ``{Segmentation of the foveal
  microvasculature using deep learning networks},'' \emph{Journal of Biomedical
  Optics}, vol.~21, no.~7, p. 075008, 2016.

\bibitem{Mou2019}
L.~Mou, Y.~Zhao, L.~Chen, J.~Cheng, Z.~Gu, H.~Hao, H.~Qi, Y.~Zheng, A.~Frangi,
  and J.~Liu, \emph{CS-Net: Channel and Spatial Attention Network for
  Curvilinear Structure Segmentation}, 10 2019, pp. 721--730.

\bibitem{AmatRoldan2015}
I.~Amat-Roldan, A.~Berzigotti, R.~Gilabert, and J.~Bosch, ``{Assessment of
  hepatic vascular network connectivity with automated graph analysis of
  dynamic contrast-enhanced us to evaluate portal hypertension in patients with
  cirrhosis: A pilot study1},'' \emph{Radiology}, vol. 277, no.~1, pp.
  268--276, 2015.

\bibitem{Alves2018}
A.~P. Alves, O.~N. Mesquita, J.~G{\'{o}}mez-Garde{\~{n}}es, and U.~Agero,
  ``{Graph analysis of cell clusters forming vascular networks},'' \emph{Royal
  Society Open Science}, vol.~5, no.~3, 2018.

\bibitem{Ritchie2013}
C.~W. Ritchie, K.~Wells, and K.~Ritchie, ``{The PREVENT research programme-A
  novel research programme to identify and manage midlife risk for dementia:
  The conceptual framework},'' \emph{International Review of Psychiatry},
  vol.~25, no.~6, pp. 748--754, 2013.

\bibitem{Jiang2018}
H.~Jiang, Y.~Wei, S.~Yingying, C.~B. Wright, X.~Sun, G.~Gregori, F.~Zheng,
  E.~A. Vanner, B.~L. Lam, T.~Rundek, and J.~Wang, ``Altered macular
  microvasculature in mild cognitive impairment and alzheimer disease,''
  \emph{Journal of neuro-ophthalmology}, vol.~38, no.~3, pp. 1536--5166, 2018.

\bibitem{VanDeKreeke2019}
J.~A. {Van De Kreeke}, H.~T. Nguyen, E.~Konijnenberg, J.~Tomassen, A.~{Den
  Braber}, M.~{Ten Kate}, M.~Yaqub, B.~{Van Berckel}, A.~A. Lammertsma, D.~I.
  Boomsma, S.~H. Tan, F.~Verbraak, and P.~J. Visser, ``{Optical coherence
  tomography angiography in preclinical Alzheimer's disease},'' \emph{British
  Journal of Ophthalmology}, pp. 157--161, 2019.

\bibitem{Spaide}
F.~J. G. W. N.~K. Spaide, Richard~F, ``Image artifacts in optical coherence
  tomography angiography,'' \emph{RETINA}, vol.~35, no.~11, p. 2163‐2180,
  2015.

\bibitem{ITK-SNAP}
P.~A. Yushkevich, J.~Piven, H.~Cody~Hazlett, R.~Gimpel~Smith, S.~Ho, J.~C. Gee,
  and G.~Gerig, ``User-guided {3D} active contour segmentation of anatomical
  structures: Significantly improved efficiency and reliability,''
  \emph{Neuroimage}, vol.~31, no.~3, pp. 1116--1128, 2006.

\bibitem{Eladawi2017}
N.~Eladawi, M.~Elmogy, O.~Helmy, A.~Aboelfetouh, A.~Riad, H.~Sandhu, S.~Schaal,
  and A.~El-Baz, ``{Automatic blood vessels segmentation based on different
  retinal maps from OCTA scans},'' \emph{Computers in Biology and Medicine},
  vol.~89, no. August, pp. 150--161, 2017.

\bibitem{Frangi1998}
A.~F. Frangi, W.~J. Niessen, K.~L. Vincken, and M.~A. Viergever, ``{Multiscale
  vessel enhancement filtering},'' in \emph{Medical Image Computing and
  Computer-Assisted Intervention - MICCAI'98}, ser. Lecture Notes in Computer
  Science, A.~C. W.M.~Wells and S.~Delp, Eds., vol. 1496.\hskip 1em plus 0.5em
  minus 0.4em\relax Berlin, Germany: Springer Verlag, 1998, pp. 130--137.

\bibitem{Soares2006}
J.~V. Soares, J.~J. Leandro, R.~M. Cesar, H.~F. Jelinek, and M.~J. Cree,
  ``{Retinal vessel segmentation using the 2-D Gabor wavelet and supervised
  classification},'' \emph{IEEE Transactions on Medical Imaging}, vol.~25,
  no.~9, pp. 1214--1222, 2006.

\bibitem{Annunziata2016}
R.~Annunziata and E.~Trucco, ``{Accelerating Convolutional Sparse Coding for
  Curvilinear Structures Segmentation by Refining SCIRD-TS Filter Banks},''
  \emph{IEEE Transactions on Medical Imaging}, vol.~35, no.~11, pp. 2381--2392,
  2016.

\bibitem{Law2008}
M.~W. Law and A.~C. Chung, ``{Three dimensional curvilinear structure detection
  using optimally oriented flux},'' in \emph{Computer Vision - ECCV 2008}, vol.
  5305.\hskip 1em plus 0.5em minus 0.4em\relax Springer, 2008, pp. 368--382.

\bibitem{Bradley}
D.~Bradley and G.~Roth, ``Adaptive thresholding using the integral image,''
  \emph{J. Graphics Tools}, vol.~12, pp. 13--21, 01 2007.

\bibitem{Li2017}
A.~Li, J.~You, C.~Du, and Y.~Pan, ``{Automated segmentation and quantification
  of OCT angiography for tracking angiogenesis progression},'' \emph{Biomedical
  Optics Express}, vol.~8, no.~12, p. 5604, 2017.

\bibitem{Otsu}
N.~{Otsu}, ``A threshold selection method from gray-level histograms,''
  \emph{IEEE Transactions on Systems, Man, and Cybernetics}, vol.~9, no.~1, pp.
  62--66, 1979.

\bibitem{Rodrigues2013}
P.~Rodrigues, P.~Guimar{\~{a}}es, T.~Santos, S.~Sim{\~{a}}o, T.~Miranda,
  P.~Serranho, and R.~Bernardes, ``{Two-dimensional segmentation of the retinal
  vascular network from optical coherence tomography},'' \emph{Journal of
  Biomedical Optics}, vol.~18, no.~12, p. 126011, 2013.

\bibitem{Kappa}
M.~L. McHugh, ``Interrater reliability: the kappa statistic,'' \emph{Biochemia
  medica}, vol.~3, pp. 276--82, 05 2012.

\bibitem{Gegundez-Arias2012}
M.~E. Gegundez-Arias, A.~Aquino, J.~M. Bravo, and D.~Marin, ``{A function for
  quality evaluation of retinal vessel segmentations},'' \emph{IEEE
  Transactions on Medical Imaging}, vol.~31, no.~2, pp. 231--239, 2012.

\bibitem{Hu2019a}
X.~Hu, L.~Fuxin, D.~Samaras, and C.~Chen, ``{Topology-Preserving Deep Image
  Segmentation},'' no. NeurIPS, pp. 1--12, 2019.

\bibitem{Schneider}
S.~Schneider and I.~F. Sbalzarini, ``{Finding faces in a planar embedding of a
  graph}.''

\bibitem{gudhi:urm}
\BIBentryALTinterwordspacing
{The GUDHI Project}, \emph{{GUDHI} User and Reference Manual},
  {3.1.1}~ed.\hskip 1em plus 0.5em minus 0.4em\relax {GUDHI Editorial Board},
  2020. [Online]. Available: \url{https://gudhi.inria.fr/doc/3.1.1/}
\BIBentrySTDinterwordspacing

\end{thebibliography}

\newpage
\appendix
\section*{Supplementary material}

\begin{figure}[!htb]
\centering
\includegraphics[width=9cm]{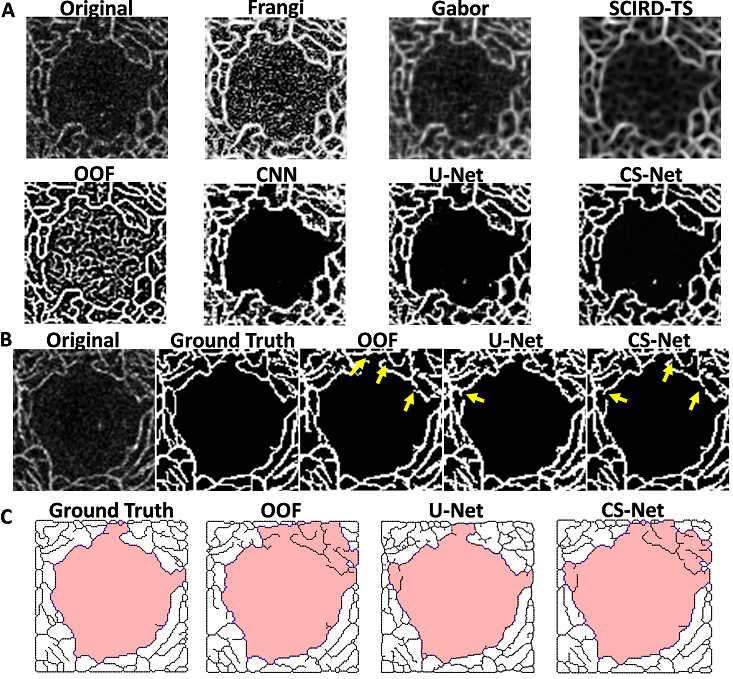}
\renewcommand{\thefigure}{S1}
\caption{Foveal images. (A) Enhancement in the foveal avascular zone. (B)  Foveal regions after binarisation: disconnections (yellow arrow)  in OOF, U-Net, and CS-Net.(C) Different FAZ detections (red) due to disconnections in the binarisation.} \label{fovea}
\end{figure}
\begin{table}[!htb]
\renewcommand{\arraystretch}{1.3}
\centering
\renewcommand{\thetable}{S1}\caption {Table of parameters for the handcrafted filters }\label{tab: params}
\begin{tabular}{ l|l|l }

\hline
\multicolumn{3}{ c }{Parameters} \\
\hline
\multirow{4}{*}{Frangi} & FrangiScaleRange  & [0.5, 2] \\
 & FrangiScaleRatio & 0.5 \\
 & FrangiBetaOne & 1 \\
 & FrangiBetaTwo & 15 \\\hline
 \multirow{3}{*}{Gabor} & scales  & [1,2,3,4] \\
 & epsilon & 4 \\
 & k0 & [0 3] \\\hline
\multirow{9}{*}{SCIRD--TS} & $fb\_parameters.sigma\_1$ & [1,5] \\
 & $fb\_parameters.sigma\_1\_step$ & 0.5 \\
 & $fb\_parameters.sigma\_2$ & [1,2] \\
 & $fb\_parameters.sigma\_2\_step$ & 0.5 \\ 
 & $fb\_parameters.k$  & [-0.1, 0.1] \\
 & $fb\_parameters.k\_step$ & 0.025 \\
 & $fb\_parameters.angle\_step$ & 10 \\
 & $fb\_parameters.filter\_size$ & 9 \\
 & $alpha$ &  0.05 \\\hline
 \multirow{3}{*}{OOF} & range  & [0.5, 2] \\
 & sigma  & 0.5 \\
 & upthreshold & 70 \\\hline
\end{tabular}

\end{table}

\begin{table}[!htb]
    \centering \renewcommand{\thetable}{S2}\caption{CNN layers architecture}  \label{tab:CNN}
  \renewcommand{\arraystretch}{1.5}
 \begin{tabular}{cccc}
 \hline
Layer  & Type   & Maps and size &  Kernel size               \\
\hline
 \centering
 0 & Input & 1 map of $61 \times 61$ neurons &      \\
 1 &    Convolution2D &  32 maps of $56 \times 56$ neurons &    $6 \times 6$ \\
 2 &    Maxpooling2D &  32 maps of $28 \times 28$ neurons &    $2 \times 2$ \\
 3 &    Convolution2D &  32 maps of $24 \times 24$ neurons &    $5 \times 5$ \\
 4 &    Maxpooling2D &  32 maps of $12 \times 12$ neurons &    $2 \times 2$ \\
 5 &    Convolution2D &  32 maps of $9 \times 9$ neurons &    $4 \times 4$ \\
 6 &    Maxpooling2D &  32 maps of $5 \times 5$ neurons &    $2 \times 2$ \\
 7 &    Dense & $150$ neurons &   \\
 8 & Dropout & & \\
 9 & Dense & 1 neuron & \\
\hline
    \end{tabular}

\end{table}

\begin{table}[!htb]
    \centering \renewcommand{\thetable}{S3}\caption{U-Net layers architecture}  \label{tab:Unet}
 \renewcommand{\arraystretch}{1.3}
 \begin{tabular}{cccc}
 \hline
Layer  & Type   & Maps and size &  Kernel size               \\
\hline
 \centering
 0 & Input & 1 map of $32 \times 32$ neurons &      \\
 1 &    Convolution2D &  32 maps of $32 \times 32$ neurons &    $3 \times 3$ \\
 2 &    Convolution2D &  32 maps of $32 \times 32$ neurons &    $3 \times 3$ \\
 3 &    Maxpooling2D &  32 maps of $16 \times 16$ neurons &    $2 \times 2$ \\
 & & &\\
 4 &    Convolution2D &  64 maps of $16 \times 16$ neurons &    $3 \times 3$ \\
 5 &    Convolution2D &  64 maps of $16 \times 16$ neurons &    $3 \times 3$ \\
 6 &    Maxpooling2D &  64 maps of $8 \times 8$ neurons &    $2 \times 2$ \\
  & & &\\
 7 &    Convolution2D &  128 maps of $8 \times 8$ neurons &    $3 \times 3$ \\
 8 &    Convolution2D &  128 maps of $8 \times 8$ neurons &    $3 \times 3$ \\
  & & &\\
 9 &   Upsampling2D &  128 maps of $16 \times 16$ neurons &    $2 \times 2$ \\
 10 &  Concatenate &  192 maps of $16 \times 16$ neurons &     \\
 11 &  Convolution2D &  64 maps of $16 \times 16$ neurons &    $3 \times 3$ \\
 12 &  Convolution2D &  64 maps of $16 \times 16$ neurons &    $3 \times 3$ \\
 
   & & &\\
 9 &   Upsampling2D &  64 maps of $32 \times 32$ neurons &    $2 \times 2$ \\
 10 &  Concatenate &  96 maps of $32 \times 32$ neurons &     \\
 11 &  Convolution2D &  32 maps of $32 \times 32$ neurons &    $3 \times 3$ \\
 12 &  Convolution2D &  32 maps of $32 \times 32$ neurons &    $3 \times 3$ \\
   & & &\\
 13 &  Convolution2D &  1 map of $32 \times 32$ neurons &    $3 \times 3$ \\

\hline
    \end{tabular}

\end{table}

\end{document}